# Exploring Combinations of Ontological Features and Keywords for Text Retrieval

Tru H. Cao, Khanh C. Le and Vuong M. Ngo

vuong.cs@gmail.com

**Abstract.** Named entities have been considered and combined with keywords to enhance information retrieval performance. However, there is not yet a formal and complete model that takes into account entity names, classes, and identifiers together. Our work explores various adaptations of the traditional Vector Space Model that combine different ontological features with keywords, and in different ways. It shows better performance of the proposed models as compared to the keyword-based Lucene, and their advantages for both text retrieval and representation of documents and queries.

## 1 Introduction

Information retrieval, in general, and text retrieval[4], in particular, is not a new area but still attracts much research effort, social and industrial interests. That is because, on the one hand, it is important for searching required information, especially on the explosive WWW, and on the other hand, there are still many open problems to be solved to enhance the existing methods or to propose new models. Retrieval precision and recall could be improved by developing appropriate models, typically as similarity-based ([5], [13]), probabilistic relevance ([15]), or probabilistic inference ([16]) ones. Semantic annotation, representation, and processing of documents and queries are another way to obtain better performance ([4], [6], [8], [17]).

Traditionally, text retrieval is only based on keywords (KW) occurring in documents and queries. Later on, word similarity and relationship are exploited to represent and match better documents to a query. However, keywords alone are not adequate, because in many domains and cases named entities (NE) constitute the user intention in a query and the main content of a document. Named entities are those that can be referred to by names, such as people, organizations, and locations ([14]). They are inherently different from words, as they represent individuals while words denote general concepts, such as types, properties, and relations. If named entities are marked up in texts then, for example, one can search for, and correctly obtain, web pages about *Saigon* as a city. Whereas current search engines like Google may return any page that contains the word *Saigon*, though it is the name of a river or a university.

There are different ontological features of named entities that can be of user interest and expressed in a query. First, the user may want to search for documents about exactly identified named entities, like the *Saigon City* in Viet Nam but not a city of the same name elsewhere. Second is the case when only the name and class of entities are of concern or available, as in searching for documents about people named *McCarthy*. Third, one may be interested in documents about entities of a certain class, like city capitals. Fourth, it is not uncommon that only entity names are the criterion of a search. In short, the possible distinct features of named entities in question are names, classes, joins of names and classes, and identifiers. Nevertheless, usually, a query cannot be completely specified without keywords, like "*economic growth of East Asian countries*", where *East Asian countries* represents named entities while *economic* and *growth* are keywords.

---

[4] In this paper we use the terms *information retrieval*, *text retrieval*, and *document retrieval* interchangeably, though they are not quite the same.

Until now, to our knowledge, there is no information retrieval model that formally integrates and treats all above-mentioned named entity features in combination with keywords. Our work presented in this paper is to explore and analyse possible combinations of ontological features and keywords in the formal framework of the Vector Space Model (VSM) and its adaptation. Implementation and experiments are also carried out to evaluate and compare the performance of developed models themselves and to the traditional purely keyword-based VSM. Section 2 recalls the basic notion of the traditional VSM and system, and its adaptation for the named entity spaces. Section 3 presents alternative adapted VSMs that combine both named entities and keywords. Section 4 is for evaluation and discussion on experimental results. In Section 5, we review related works in comparison with our approach. Finally, Section 6 gives some concluding remarks.

## 2   Ontology-Based Multi-Vector Space Models

Despite having known disadvantages, VSM is still a popular model and a basis to develop other models for information retrieval, because it is simple, fast, and its ranking method is in general either better or almost as good as a large variety of alternatives ([1]). We recall that, in the keyword-based VSM, each document is represented by a vector over the space of keywords of discourse. Conventionally, the weight corresponding to a term dimension of the vector is a function of the occurrence frequency of that term in the document, called *tf*, and the inverse occurrence frequency of the term across all the existing documents, called *idf*. The similarity degree between a document and a query is then defined as the cosine of their representing vectors.

Given a query, the retrieval process composes of two main stages, namely, document filtering and document ranking. The former selects those documents that satisfy the Boolean expression of keywords as specified in the query. For example, if the query is $k_1 \vee k_2$, and $D_1$ and $D_2$ are respectively the sets of documents that contain $k_1$ and $k_2$, then $D_1 \cup D_2$ is the set of selected documents. In the latter, those selected documents are ranked by their similarity degrees to the query as calculated above.

With terms being keywords, the traditional VSM cannot satisfactorily represent the semantics of texts with respect to the named entities they contain, such as for the following queries:

$Q_1$: Search for documents about *cities*.
$Q_2$: Search for documents about *Saigon City*.
$Q_3$: Search for documents about *Hanoi Tower*.
$Q_4$: Search for documents about *Hanoi University of Technology*.

That is because, for $Q_1$, a target document does not necessarily contain the keyword *city*, but only some named entities of the class *City*, i.e., real cities in the world. For $Q_2$, a target document may mention about *Saigon City* by other names, i.e., the city's aliases, such as *Ho Chi Minh City*. On the other hand, documents containing entities named *Saigon* but not being cities, like *Saigon River*, are not target documents. For $Q_3$, documents about *Hanoi* as a city or a university are not target documents at all, though containing the keyword *Hanoi*. Meanwhile, $Q_4$ targets at documents about a precisely identified named entity, i.e., *Hanoi University of Technology*, not other universities of similar names. Therefore, simple keyword looking up and matching may fail to give expected answers.

For formally representing documents (and queries) by named entity features, we define the triple ($N$, $C$, $I$) where $N$, $C$, and $I$ are respectively the sets of names, classes, and identifiers of named entities in the ontology of discourse. Then:

1. Each document $d$ is modelled as a subset of $(N \cup \{*\}) \times (C \cup \{*\}) \times (I \cup \{*\})$, where '*' denotes an unspecified name, class, or identifier of a named entity in $d$, and
2. $d$ is represented by the quadruple ($\vec{d}_N$, $\vec{d}_C$, $\vec{d}_{NC}$, $\vec{d}_I$), where $\vec{d}_N$, $\vec{d}_C$, $\vec{d}_{NC}$, and $\vec{d}_I$ are respectively vectors over $N$, $C$, $N \times C$, and $I$.

A feature of a named entity could be unspecified due to the user intention expressed in a query, the incomplete information about that named entity in a document, or the inability of an employed NE recognition engine to fully recognize it. Each of the four component vectors introduced above for a document can be defined as a vector in the traditional *tf.idf* model on the corresponding space of entity names, classes, name-class pairs, or identifiers, instead of keywords. However, there are two following important differences with those ontological features of named entities in calculation of their frequencies:

1. The frequency of a name also counts identical entity aliases. That is, if a document contains an entity having an alias identical to that name, then it is assumed as if the name occurred in the document. For example, if a document refers to *Saigon City*, then each occurrence of that entity in the document is counted as one occurrence of the name *Ho Chi Minh City*, because it is an alias of *Saigon City*.
2. The frequency of a class also counts occurrences of its subclasses. That is, if a document contains an entity whose class is a subclass of that class, then it is assumed as if the class occurred in the document. For example, if a document refers to *Saigon City*, then each occurrence of that entity in the document is counted as one occurrence of the class *Location*, because *City* is a subclass of *Location*.

The similarity degree of a document $d$ and a query $q$ is then defined to be, where $w_N + w_C + w_{NC} + w_I = 1$:

$$sim(\vec{d}, \vec{q}) = w_N.cosine(\vec{d}_N, \vec{q}_N) + w_C.cosine(\vec{d}_C, \vec{q}_C) + w_{NC}.cosine(\vec{d}_{NC}, \vec{q}_{NC}) + w_I.cosine(\vec{d}_I, \vec{q}_I) \quad \text{(Eq. 1)}$$

We deliberately leave the weights in the sum unspecified, to be flexibly adjusted in applications, depending on user-defined relative significances of the four ontological features. We note that the join of $\vec{d}_N$ and $\vec{d}_C$ cannot replace $\vec{d}_{NC}$ because the latter is concerned with entities of certain name-class pairs. Meanwhile, $\vec{d}_{NC}$ cannot replace $\vec{d}_I$ because there may be different entities of the same name and class (e.g. there are different cities named *Moscow* in the world). Also, since names and classes of an entity are derivable from its identifier, products of *I* with *N* or *C* are not included. In brief, here we generalize the notion of terms being keywords in the traditional VSM to be entity names, classes, name-class pairs, or identifiers, and use four vectors on those spaces to represent a document or a query for text retrieval.

There are still possible variations of this proposed ontology-based multi-vector space model that are worth exploring. Firstly, that is due to overlapping of those four types of generalized terms in a query, which all convey information about the documents that a user wants to search for. For example, given a query containing *Ho Chi Minh City*, this entity includes all the four terms, namely the identifier of the entity itself, the name-class pair (*Ho Chi Minh*, *City*), the class *City*, and the name *Ho Chi Minh*. We call these variations overlapped or non-overlapped models, respectively denoted by NEo or NEn, depending on whether term overlapping is taken into account or not. Figure 2.1 shows a query in the TIME test collection (available with [2]) and its corresponding sets of ontological terms that we extract for the two models, where *InternationalOrganization_T.17* is the identifier of *United Nations* in the knowledge base of discourse.

---

Query: "*Countries have newly joined the United Nations*".

Overlapped ontological term set:
{(*/*Country*/*), (*United Nations*/*/*), (*/*InternationalOrganization*/*), (*United Nations*/
 *InternationalOrganization*/*), (*United Nations*/ *InternationalOrganization*/*InternationalOrganization_T.17* )}

Non-overlapped ontological term set:
{(*/*Country*/*), (*United Nations*/*InternationalOrganization*/*InternationalOrganization_T.17*)}

---

**Fig. 2.1.** Overlapped and non-overlapped ontological terms extracted from a query

As in the traditional VSM retrieval process, after the Boolean document filtering stage, let $D_N$, $D_C$, $D_{NC}$, and $D_I$ be the respective sets of obtained documents containing generalized terms of the four ontological features in a query. For the document ranking stage, we take the intersection of $D_N$, $D_C$, $D_{NC}$, and $D_I$ in the overlapped model or their union in the non-overlapped model, respectively, as the set of documents to be ranked and returned for the query. This application of intersection or union operations can be justified as responding to the overlapping effect, which is supported by experimental results shown later.

## 3 Combining Named Entities and Keywords

Clearly, named entities alone are not adequate to represent a text. For example, in the query in Figure 2.1, *joined* is a keyword to be taken into account, and so are *Countries* and *United Nations*, which can be concurrently treated as both keywords and named entities. Therefore, a document can be represented by one vector on keywords and four vectors on ontological terms. Then, given a query, after the document filtering stage, one can take either the intersection or the union of the document set satisfying the Boolean expression of the keywords and the document set satisfying the Boolean expression of the named entities in the query.

Regarding also overlapping or non-overlapping of ontological terms as discussed in Section 2, one have four alternative models combining keywords and named entities, denoted by KW∩NEo, KW∩NEn, KW∪NEo, and KW∪NEn. The similarity degree of a document *d* and a query *q* is then defined as follows, where $w_N + w_C + w_{NC} + w_I = 1$, $\alpha \in [0, 1]$, and $\vec{d}_{KW}$ and $\vec{q}_{KW}$ are respectively the vectors representing the keyword features of *d* and *q*:

$$sim(\vec{d}, \vec{q}) = \alpha.[w_N.cosine(\vec{d}_N, \vec{q}_N) + w_C.cosine(\vec{d}_C, \vec{q}_C) + w_{NC}.cosine(\vec{d}_{NC}, \vec{q}_{NC}) + w_I.cosine(\vec{d}_I, \vec{q}_I)] + (1 - \alpha).cosine(\vec{d}_{KW}, \vec{q}_{KW})$$ (Eq. 2)

We now explore another adapted VSM that combines keywords and named entities. That is we unify and treat all of them as generalized terms, where a term is counted either as a keyword or a named entity but not both. Each document is then represented by a single vector over that generalized term space. Document vector representation, filtering, and ranking are performed as in the traditional VSM, except for taking into account entity aliases and class subsumption as presented in Section 2. We denote this model by KW+NE. Figure 3.1 show another query in the TIME test collection and its corresponding key term sets for the multi-vector space models and the KW+NE model.

---

Query: "*U.N. team survey of public opinion in North Borneo and Sarawak on the question of joining the federation of Malaysia*".

Multi-vector space models (KW∩NEo, KW∩NEn, KW∪NEo, KW∪NEn):
Keywords = {*U.N, opinion, North Borneo, Sarawak, join, federation, Malaysia*}
Onto-terms = {(*U.N./InternationalOrganization/InternationalOrganization_T.17*), (*North Borneo/Province/ Province_T.2189*), (*Sarawak/Location/\**), (*Malaysia/Country/Country_T.MY*)}

KW+NE model:
Generalized terms = {(*U.N./InternationalOrganization/InternationalOrganization_T.17*), *opinion*, (*North Borneo/Province/Province_T.2189*), (*Sarawak/Location/\**), *join*, *federation*, (*Malaysia/Country/Country_T.MY*)}

---

**Fig. 3.1.** Keywords, ontological terms, and generalized terms extracted from a query

## 4  Implementation and Experimentation

We have implemented the above-adapted VSMs by employing and modifying Lucene, a general VSM-based open source for storing, indexing and searching documents ([7]). We have evaluated and compared the new models in terms of precision-recall (P-R) curves and single F-measure values. For each query in a test collection, we adopt the common method in [11] to obtain the corresponding P-R curve. That is, the returned documents are examined from the top to the bottom, regarding their similarity degrees to the query. At each step, the precision and recall for the documents that have been examined are calculated, creating one point of the curve.

In order to obtain the average P-R curve over all the queries in the test collection, each query curve is interpolated to the eleven standard recall levels that are 0%, 10%, …, 100%, as in [1]. The interpolated precision for the *i*-th query at the *j*-th standard recall level $r_j$ ($j \in \{0, 1, …, 10\}$) is defined by $P_i(r_j) = max_{r_j \leq r \leq r_{j+1}} P_i(r)$. Given $N_q$ as the number of queries, the average precision at $r_j$ over all the queries is then computed by $\overline{P}(r_j) = \sum_{i=1}^{N_q} \frac{P_i(r_j)}{N_q}$. Consequently, the interpolated F-measure value for the *i*-th query at $r_j$ is $F_i(r_j) = \dfrac{2.P(r_j).r_j}{P(r_j) + r_j}$, and the average F-measure value at $r_j$ over all the queries is $\overline{F}(r_j) = \sum_{i=1}^{N_q} \dfrac{F_i(r_j)}{N_q}$.

We have conducted experimentation on the TIME collection, containing 425 documents and 83 queries. The ontology and NE recognition engine of KIM ([10]) are employed to automatically annotate named entities in documents. For the queries, we manually extract and mark their named entities and keywords, to represent their meanings concisely and appropriately for document retrieval. In the experiments, we set the weights $w_N = w_C = w_{NC} = w_I = 0.25$ and $\alpha = 0.5$, assuming that the keyword and named entity dimensions are of equal importance.

Table 4.1 presents the average precisions of the keyword-based VSM by Lucene itself, the NE-based overlapped/non-overlapped models, and the KW-NE-based[5] models combining named entities and keywords, at each of the standard recall levels. Table 4.2 shows their average F-measure values. One can observe that, for all the models, the maximum F-measure values are achieved at the 50% recall level. The performances of the NEo and NEn models are quite similar (39.1 and 38.9), so are those of the KW-NE-based models (around 42.0). Therefore, we take the NEn model and the KW+NE model as representatives of these two groups, respectively. The similar performances of the models in each group justify our use of intersection or union on filtered document sets in accordance to overlapping or non-overlapping application on query terms.

**Table 4.1.** The average precisions at the eleven standard recall levels

| | Recall (%) | | | | | | | | | | | |
|---|---|---|---|---|---|---|---|---|---|---|---|---|
| | 0 | 10 | 20 | 30 | 40 | 50 | 60 | 70 | 80 | 90 | 100 | |
| Lucene | 56.3 | 56.52 | 55.27 | 53.12 | 50.98 | 47.62 | 27.64 | 21.25 | 18.74 | 8.076 | 5.28 | Precision (%) |
| NEo | 54.5 | 53.4 | 52.48 | 49.57 | 49.25 | 48.01 | 29.83 | 22.75 | 20.38 | 11.81 | 11.18 | |
| NEn | 54.86 | 53.55 | 53.6 | 50.64 | 49.35 | 48.51 | 28.41 | 22.65 | 19.53 | 8.95 | 8.183 | |
| KW+NE | 62.39 | 61.95 | 61.14 | 59.35 | 57.8 | 56.24 | 31.95 | 24.45 | 21.45 | 8.366 | 5.711 | |
| KW∪NEo | 60.58 | 60.36 | 59.46 | 56.35 | 56.27 | 55.64 | 32.78 | 22.41 | 20.02 | 7.925 | 4.976 | |
| KW∩NEo | 60.43 | 60.21 | 59.32 | 56.2 | 56.13 | 55.5 | 34.37 | 24.91 | 22.03 | 13.14 | 11.97 | |
| KW∪NEn | 60.8 | 60.58 | 60.23 | 56.7 | 56.42 | 55.28 | 32.33 | 22.72 | 18.87 | 7.101 | 4.617 | |
| KW∩NEn | 60.81 | 60.59 | 60.24 | 56.71 | 56.44 | 55.41 | 33.68 | 24.43 | 20.29 | 10.48 | 9.37 | |

**Table 4.2.** The average F-measure values at the eleven standard recall levels

| | Recall (%) | | | | | | | | | | | |
|---|---|---|---|---|---|---|---|---|---|---|---|---|
| | 0 | 10 | 20 | 30 | 40 | 50 | 60 | 70 | 80 | 90 | 100 | |
| Lucene | 0 | 12.68 | 21.97 | 29.3 | 34.86 | **37.93** | 26.25 | 22.8 | 21.77 | 11.26 | 8.532 | F-measure (%) |
| NEo | 0 | 13.13 | 22.45 | 29.22 | 34.99 | **39.1** | 29.44 | 25.46 | 24.29 | 17.04 | 16.82 | |
| NEn | 0 | 13.13 | 22.58 | 29 | 34.58 | **38.9** | 27.57 | 24.21 | 22.68 | 13.27 | 12.76 | |
| KW+NE | 0 | 13.3 | 23.43 | 31.47 | 37.67 | **42.46** | 28.83 | 24.77 | 23.55 | 11.84 | 9.189 | |
| KW∪NEo | 0 | 13.08 | 22.97 | 30.41 | 36.76 | **41.89** | 28.83 | 23.06 | 22.03 | 10.93 | 7.951 | |
| KW∩NEo | 0 | 13.61 | 23.63 | 31.08 | 37.43 | **42.51** | 32 | 27.5 | 26.11 | 18.46 | 18.01 | |
| KW∪NEn | 0 | 13.1 | 23.12 | 30.42 | 36.78 | **41.86** | 28.63 | 22.92 | 20.98 | 9.877 | 7.311 | |
| KW∩NEn | 0 | 13.5 | 23.63 | 30.97 | 37.37 | **42.53** | 31.28 | 26.47 | 24.28 | 15.28 | 14.68 | |

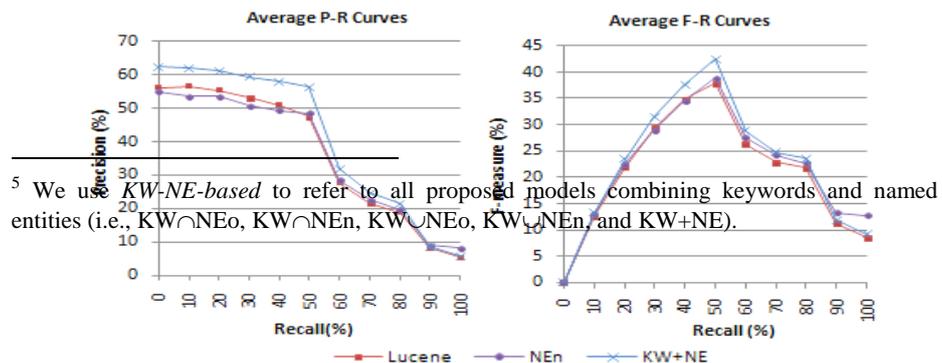

---

[5] We use *KW-NE-based* to refer to all proposed models combining keywords and named entities (i.e., KW∩NEo, KW∩NEn, KW∪NEo, KW∪NEn, and KW+NE).

**Fig. 4.1.** Average P-R and F-R curves of Lucene, NEn, and KW+NE models

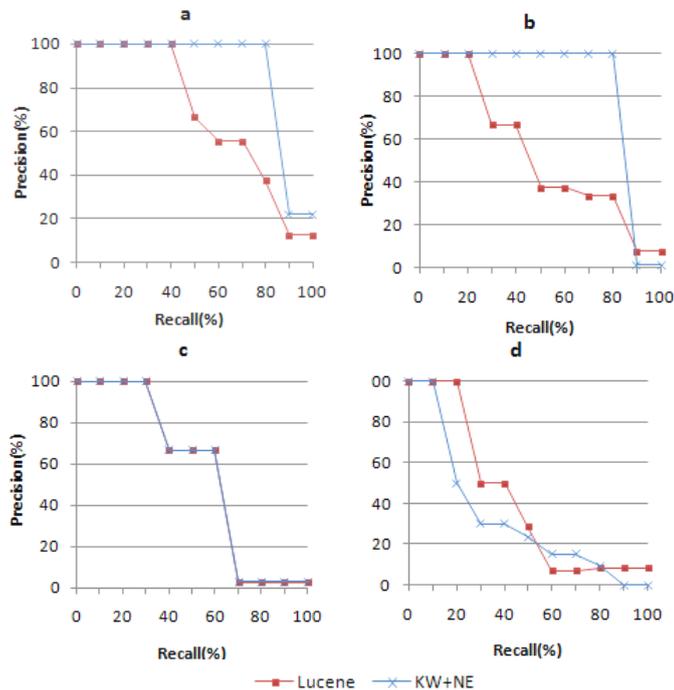

**Fig. 4.2.** Performances on typical queries of Lucene and KW+NE models

Overall, KW+NE is better than NEn (42.46 versus 38.9), and both are better than the Lucene baseline (37.93). The difference would be larger on a test collection involving more named entities and ontological terms than in the TIME one. Figure 4.1 illustrates the average P-R and average F-R curves of the three models. We have also examined some typical queries for which KW+NE is better than, as good as, or worse than Lucene, as shown in Figure 4.2. Following are those queries and our analysis.

**Query a.** "*Kennedy administration pressure on Ngo Dinh Diem to stop suppressing the buddhists*". For this query, our single vector model KW+NE performs better than Lucene because the latter fails to recognize aliases of the named entities in the query. There are two ontological terms here, namely, (*Kennedy/Person/\**) and (*Ngo Dinh Diem/Person/\**). In the document collection, *Kennedy* also occurs as *John Kennedy*, and *Ngo Dinh Diem* has two other aliases *NgoDinh Diem* and *Diem*. These aliases are used frequently in the documents, but keyword-based search sees them as different terms, which leads to a reduction in retrieval precision.

**Query b.** "*Persons involved in the Viet Nam war*". For this query, KW+NE also outperforms Lucene. That can be explained by the fact that, while keyword-based search looks for documents explicitly containing the words *person* or *persons*, KW+NE recognizes and selects also those documents that contain named entities of the class *Person*. It boosts up the ranking values of relevant documents to be placed at the top of the returned document list.

**Query c.** "*Somalia is involved in border disputes with its neighbors what military aid is being supplied to Somalia by Russia*". This is a case when KW+NE and Lucene have no performance difference. That is because there are no aliases of *Somalia* and *Russia* in the document collection. So, what actually happens is that KW+NE matches identifiers with identifiers whereas Lucene matches names with names, representing the two named entities. Without aliases, that obviously does not affect the results.

**Query d.** "*Indian fears of another Chinese invasion*". For this query, Lucene performs slightly better than KW+NE. Here, two implicit named entities *India* and *China* are manually extracted from *Indian* and *Chinese*, respectively. However, KIM NE recognition engine could not detect named entities implicitly occurring in a document under the adjective form. So, with the KW+NE model, a document just containing the keywords *Indian* and *Chinese* is not considered as relevant to the query, while with Lucene they are. That explains the difference.

We note that the performance of any system relying on named entities to solve a particular problem partly depends on that of the NE recognition module in a preceding stage. However, in research for models or methods, the two problems should be separated. This paper is not about NE recognition and our experiments incur errors of the employed KIM engine, whose current average precision and recall are respectively 90% and 86%.

Among the KW-NE-based models, the KW+NE model is straightforward and simple, unifying keywords and named entities as generalized terms, while having comparable performance as the others. Meanwhile, the multi-vector models can be useful for clustering documents into a hierarchy via top-down phases each of which uses one of the four NE-based vectors presented above (cf. [3]). For example, given a set of geographical documents, one can first cluster them into groups of documents about rivers and mountains, i.e., clustering with respect to entity classes. Then, the documents in the river group can be clustered further into subgroups each of which is about a particular river, i.e., clustering with respect to entity identifiers. As another example of combination of clustering objectives, one can first make a group of documents about entities named *Saigon*, by clustering them with respect to entity names. Then, the documents within this group can be clustered further into subgroups for *Saigon City*, *Saigon River*, and *Saigon Market*, for instance, by clustering them with respect to entity classes. Another advantage of splitting document representation into four component vectors is that, searching and matching need to be performed only for those components that are relevant to a certain query.

## 5 Related Works

In [12], a probabilistic relevance model was introduced for searching passages about certain biomedical entity types (i.e., classes) only, such as genes, diseases, or drugs. Also in the biomedical domain, the similarity-based model in [18] considered concepts being genes and medical subject headings, such as *purification*, *HNF4*, or *hepatitis B virus*. Concept synonyms, hypernyms, and hyponyms were taken into account, which respectively corresponded to entity aliases, super-classes, and subclasses in our NE-based models. A document or query was represented by two component vectors, one of which was for concepts and the other for words. A document was defined as being more similar to a query than another document if the concept component of the former is closer to that of the query. If the two concept components were equally similar to that of the query, then the similarity between the word components of the two documents and that of the query would decide. However, as such, the word component was treated as only secondary in the model, and its domain was just limited within biomedicine. Recently, [9] researched and showed that NE normalization improved retrieval performance. The work however considered only entity names and that normalization issue was in fact what we call aliasing here.

Two closely related works to ours are [4] and [6]. In [4], the authors adapted the traditional VSM with vectors over the space of NE identifiers in the knowledge base of discourse. For each document or query, the authors also applied a linear combination of its NE-identifier-based vector and keyword-based vector with the equal weights of 0.5. The system was tested on the authors' own dataset. The main drawback was that every query had to be posed using RDQL, a query language for RDF, to first look up in the system's knowledge base those named entities that satisfied the query, before its vector could be constructed. For example, given the query searching for documents about *Basketball Player*, its vector would be defined by the basketball players identified in the knowledge base. This step of retrieving NE identifiers was unnecessarily time consuming. Moreover, any knowledge base is usually incomplete, so documents containing certain basketball players not existing in the knowledge base would not be returned. In our proposed models, the query and document vectors on the entity class *Basketball Player* can be constructed and matched right away.

Meanwhile, the LRD (Latent Relation Discovery) model proposed in [6] used both keywords and named entities as terms for a single vector space. The essential of the model was that it enhanced the content description of a document by those terms that did not exist, but were related to existing terms, in the document. The relation strength between terms was based on their co-occurrence. The authors tested the model on 20 randomly chosen queries from 112 queries of the CISI dataset ([2]), achieving the maximum F-measure of 19.3. That low value might be due to the dataset containing few named entities. Anyway, the model's drawback as compared to our KW+NE model is that it used only entity names but not all ontological features. Consequently, it cannot support queries searching for documents about entities of particular classes, name-class pairs, or identifiers.

## 6 Conclusion

We have presented various adapted VSMs that take into account possible combinations of ontological features with keywords, which all yield nearly the same performance and are better than the keyword-based Lucene. Our consideration of entity name aliases and class subsumption is logically sound and empirically verified. We have shown that overlapping of ontological features if applied to a query can be compensated by taking intersection of the selected document sets with respect to each of the features. Also, retrieval performance is not sensitive to the choice of intersection or union of the selected documents satisfying the keyword expression and that for the named entity expression in the query.

For its uniformity and simplicity, we propose the single vector KW+NE model for text retrieval. Meanwhile, the multi-vector model is useful for document clustering with respect to various ontological features. These are the first basic models that formally accommodate all entity names, classes, joint names and classes, and identifiers. Within the scope of this paper, we have not considered similarity and relatedness of generalized terms of keywords and named entities. This is currently under our investigation expected to increase the overall performance of the proposed models.

## References

1. Baeza-Yates, R., Ribeiro-Neto, B. 1999: Modern Information Retrieval. Addison-Wesley.
2. Buckley, C. 1985: Implementation of the SMART Information Retrieval System. Technical Report 85-686, Cornell University.
3. Cao, T.H., Do, H.T., Hong, D.T., Quan, T.T. 2008: Fuzzy Named Entity-Based Document Clustering. In: Proceedings of the 17th IEEE International Conference on Fuzzy Systems, pp. 2028-2034.
4. Castells, P., Vallet, D., Fernández, M.: An Adaptation of the Vector Space Model for Ontology-Based Information Retrieval. IEEE Transactions of Knowledge and Data Engineering, 2007, Vol. 19, No. 2, pp. 261-272.
5. Dominich, S. 2002: Paradox-Free Formal Foundation of Vector Space Model. In: Proceedings of the ACM SIGIR 2002 Workshop on Mathematical/Formal Methods in Information Retrieval, pp. 43-48.
6. Gonçalves, A., Zhu, J., Song, D., Uren, V., Pacheco, R. 2006: LRD: Latent Relation Discovery for Vector Space Expansion and Information Retrieval. In: Proceedings of the 7th International Conference on Web-Age Information Management.
7. Gospodnetic, O. 2003: Parsing, Indexing, and Searching XML with Digester and Lucene. Journal of IBM DeveloperWorks.
8. Guha, R., McCool, R., Miller, E. 2003: Semantic Search. In: Proceedings of the 12th International Conference on World Wide Web, pp. 700-709.
9. Khalid, M.A., Jijkoun, V., de Rijke, M. 2008: The Impact of Named Entity Normalization on Information Retrieval for Question Answering. In: Proceedings of the 30th European Conference on IR Research, LNCS Vol. 4956, Springer, pp. 705-710.
10. Kiryakov, A., Popov, B., Terziev, I., Manov, D., Ognyanoff, D. 2005: Semantic Annotation, Indexing, and Retrieval. Journal of Web Semantics 2.
11. Lee, D.L., Chuang, H., Seamons, K. 1997: Document Ranking and the Vector-Space Model. IEEE Software 14, pp. 67-75.
12. Meij, E., Katrenko, S. 2007: Bootstrapping Language Associated with Biomedical Entities. In: Proceedings of the 16th Text REtrieval Conference.
13. Salton, G., Wong, A., Yang, C.S.: A Vector Space Model for Automatic Indexing. Communications of the ACM **18** (1975) 613-620.
14. Sekine, S.: Named Entity: History and Future. Proteus Project Report (2004).
15. Sparck Jones, K., Walker, S., Robertson, S.E.: A Probabilistic Model of Information Retrieval: Development and Comparative Experiments – Part 1 and Part 2. Information Processing and Management **36** (2000) 779-808 and 809-840.
16. van Rijbergen, C.J.: A Non-Classical Logic for Information Retrieval. The Computer Journal **29** (1986) 481-485.
17. Varelas, G., Voutsakis, E., Raftopoulou, P., Petrakis, E.G.M., Milios, E.E.: Semantic Similarity Methods in WordNet and Their Application to Information Retrieval on the Web. In: Proceedings of the 7th Annual ACM Intl Workshop on Web Information and Data Management (2005) 10-16.
18. Zhou, W., Yu, C.T., Torvik, V.I., Smalheiser, N.R.: A Concept-based Framework for Passage Retrieval in Genomics. In: Proceedings of the 15th Text REtrieval Conference (2006).